\documentclass[aps,twocolumn,nofootinbib,floatfix,superscriptaddress]{revtex4}
\usepackage{amsfonts}
\usepackage{amssymb}
\usepackage{amsmath}
\usepackage{graphics}
\usepackage{graphicx}
\usepackage{soul}
\usepackage{natbib}
\usepackage{bm}
\usepackage{mathtools}
\usepackage{stmaryrd}
\usepackage{epstopdf}
\usepackage{color}
\usepackage{lipsum}
\usepackage{balance}
\usepackage{enumitem}
\usepackage[colorlinks=true,citecolor=black,linkcolor=black,urlcolor=black,filecolor=black]{hyperref}

\usepackage{acronym}
\newacro{BL}{Balescu--Lenard}
\newcommand{\BL}{\ac{BL}}
\newacro{DF}{distribution function}
\newacroplural{DF}[DFs]{distribution functions}
\newcommand{\DF}{\ac{DF}}
\newcommand{\DFs}{\acp{DF}}
\newacro{HMF}{Hamiltonian Mean Field}
\newcommand{\HMF}{\ac{HMF}}

\newcommand{\rd}{\mathrm{d}}
\newcommand{\re}{\mathrm{e}}
\newcommand{\ri}{\mathrm{i}}
\newcommand{\Mtot}{M_{\mathrm{tot}}}
\newcommand{\vtheta}{\vartheta}
\newcommand{\bw}{\mathbf{w}}
\newcommand{\hbL}{\widehat{\mathbf{L}}}
\newcommand{\hbLp}{\widehat{\mathbf{L}}^{\prime}}
\newcommand{\Uext}{U_{\mathrm{ext}}}
\newcommand{\hbz}{\widehat{\mathbf{z}}}
\newcommand{\p }{\partial}
\newcommand{\bS}{\mathbf{S}}
\newcommand{\mO}{\mathcal{O}}
\newcommand{\Etot}{E_{\mathrm{tot}}}
\newcommand{\bO}{\mathbf{\Omega}}
\newcommand{\mP}{\mathcal{P}}
\newcommand{\mU}{\mathcal{U}}
\newcommand{\bJ}{\mathbf{J}}
\newcommand{\deltaD}{\delta_{\mathrm{D}}}
\newcommand{\bk}{\mathbf{k}}
\newcommand{\Td}{T_{\mathrm{d}}}
\newcommand{\oF}{\overline{F}}
\newcommand{\ot}{\overline{t}}
\newcommand{\Tr}{T_{\mathrm{r}}}
\newcommand{\FB}{F_{\mathrm{B}}}
\newcommand{\Jp}{J^{\prime}}
\newcommand{\res}{\mathrm{res}}
\newcommand{\bwp}{\mathbf{w}^{\prime}}
\newcommand{\thetap}{\theta^{\prime}}
\newcommand{\half}{\tfrac{1}{2}}
\newcommand{\third}{\tfrac{1}{3}}
\newcommand{\mF}{\mathcal{F}}
\newcommand{\kp}{k^{\prime}}
\newcommand{\mR}{\mathcal{R}}
\newcommand{\mC}{\mathcal{C}}
\newcommand{\omC}{\overline{\mathcal{C}}}
\newcommand{\omU}{\overline{\mathcal{U}}}
\newcommand{\obk}{\overline{\mathbf{k}}}
\newcommand{\ext}{\mathrm{ext}}
\newcommand{\psid}{\psi^{\mathrm{d}}}
\newcommand{\tot}{\mathrm{tot}}
\newcommand{\eps}{\epsilon}
\newcommand{\cst}{\mathrm{cst.}}
\newcommand{\omegaR}{\omega_{\mathrm{R}}}

\usepackage{soul} \usepackage{xcolor}

\begin{document}

\title{Kinetic theory of one-dimensional inhomogeneous long-range interacting\\$N$-body systems at order ${1/N^{2}}$ without collective effects}

\author{Jean-Baptiste Fouvry}
\affiliation{Institut d'Astrophysique de Paris, UMR 7095, 98 bis Boulevard Arago, F-75014 Paris, France}

\begin{abstract}
Long-range interacting systems irreversibly relax
as a result of their finite number of particles, $N$.
At order ${1/N}$, this process is described by the inhomogeneous
Balescu--Lenard equation. Yet, this equation exactly vanishes
in one-dimensional inhomogeneous systems
with a monotonic frequency profile
and sustaining only 1:1 resonances.
In the limit where collective effects can be neglected,
we derive a closed and explicit ${1/N^{2}}$ collision operator
for such systems.
We detail its properties
highlighting in particular how it satisfies an $H$-theorem
for Boltzmann entropy.
We also compare its predictions
with direct $N$-body simulations.
Finally, we exhibit a generic class of long-range interaction potentials
for which this ${ 1/N^{2}}$ collision operator exactly vanishes.
\end{abstract}
\maketitle

\section{Introduction}
\label{sec:Introduction}

Because they are composed of a finite number of particles,
long-range interacting $N$-body systems unavoidably relax
towards their thermodynamical equilibrium~\citep{Nicholson1992,BinneyTremaine2008,Bouchet+2012,Campa+2014},
should it exist~\citep{Padmanabhan1990,Chavanis2006}.
Such a dynamics is sourced
by Poisson shot noise: for a fixed total mass,
the larger the number of particles, $N$,
the slower the diffusion.
Kinetic theory aims at describing
this long-term relaxation.
Here, we are interested in inhomogeneous systems,
i.e.\ systems with a non-trivial mean-field orbital structure~\citep{BinneyTremaine2008}.
When limited to ${1/N}$ effects,
i.e.\ two-body correlations, the system's evolution
is generically described by the inhomogeneous \BL\ equation~\citep{Heyvaerts2010,Chavanis2012}.
When collective effects are neglected,
i.e.\ when one neglects the system's ability to amplify
its own self-generated fluctuations~\citep{Nelson+1999},
this intricate kinetic equation becomes the inhomogeneous
Landau equation~\citep{Chavanis2013}.
Since both equations are valid at order ${1/N}$,
they describe evolutions on timescales
of order ${ N \Td }$, with $\Td$ the dynamical time.

For ${1D}$ inhomogeneous systems
with a monotonic frequency profile
and sustaining only {1:1} resonances,
the \BL\ collision term vanishes exactly~\citep[see, e.g.\@,][]{Eldridge+1963,Dubin2003,Bouchet+2005,Chavanis+2007,Gupta+2011,Barre+2014,Lourenco+2015}.
This is a kinetic blocking
and we refer to~\cite{Fouvry+2020} for a detailed review
of the literature on that regard.
Kinetically blocked systems can only evolve under the effect
of three-body correlations.
Their relaxation
occurs therefore on a timescale of order ${N^{2}\Td}$,
or even larger.
In this paper, we derive
an appropriate kinetic equation for this (very) slow process.

Steps in that direction were successively
performed by: (i)~\cite{RochaFilho+2014}
which made a first attempt at deriving a ${1/N^{2}}$
equation for the one-dimensional (homogeneous)
\HMF\ model~\citep{Antoni+1995}
starting from the BBGKY hierarchy and neglecting
collective effects;
(ii) this was further clarified by~\cite{Fouvry+2019N2}
which emphasised the main properties of this collision operator
and compared it with numerical simulations;
(iii) finally, these results were generalised in~\cite{Fouvry+2020}
to homogeneous systems with an arbitrary interaction potential.

The kinetic equation derived in~\cite{Fouvry+2020}
was restricted to homogeneous ${1D}$ systems.
Here, we go beyond this limitation
and focus on inhomogeneous systems,
while still neglecting collective effects.
In the limit where collective amplification can be neglected,
we present a closed, explicit and well-posed kinetic equation
that describes these systems' relaxation on ${ N^{2} \Td }$ timescales.
In addition to reviewing the key properties of this collision operator,
we also quantitatively compare its predictions
with direct $N$-body simulations.
Remarkably, we present a class of interaction potentials
for which this ${1/N^{2}}$ collision term exactly vanishes
whatever the system's distribution function:
we call this a second-order kinetic blocking.

The paper is organised as follows.
In~\S\ref{sec:KineticEquation}, we spell out
the inhomogeneous ${1/N^2}$ kinetic equation,
as given by Eq.~\eqref{kin_eq}.
In~\S\ref{sec:Properties}, we present the main
properties of this equation,
while in~\S\ref{sec:Steady}, we investigate
its steady states.
In~\S\ref{sec:Application}, we quantitatively
compare the prediction of this theory
with direct numerical simulations
of particles interacting on the unit sphere.
Finally, we conclude in~\S\ref{sec:Conclusions}.
In all these sections, technical details
are either deferred to Appendices
or to appropriate references.

\section{Kinetic equation}
\label{sec:KineticEquation}

We consider a population of $N$ particles
of individual mass ${ \mu \!=\! \Mtot / N }$
with $\Mtot$ the system's total mass.
The ${1D}$ canonical (specific) phase coordinates
are denoted by ${ \bw \!=\! (\theta , J) }$
with $\theta$ the ${2\pi}$-periodic angle
and $J$ the action~\citep{BinneyTremaine2008}.
The system's total specific Hamiltonian reads
\begin{equation}
H = \sum_{i=1}^{N} \Uext (\bw_{i}) + \sum_{i < j}^{N} \mu \, U (\bw_{i} , \bw_{j}) ,
\label{Htot_init}
\end{equation}
with ${ \Uext (\bw) }$ a given external potential
and ${ U (\bw , \bwp) }$ the pairwise interaction potential
between the particles. We assume that the pairwise interaction
satisfies the symmetries ${ U (\bw , \bwp) \!=\! U \big( |\theta \!-\! \thetap| , \{ J , \Jp \} \big) }$. When Fourier expanded w.r.t.\ the angles, it becomes
\begin{equation}
U (\bw , \bwp) = \sum_{\mathclap{k = - \infty}}^{\mathclap{+ \infty}} U_{k} ( J, \Jp) \, \re^{\ri k (\theta - \thetap)} ,
\label{def_Uk}
\end{equation}
where the coefficients, ${ U_{k} (J,\Jp) \!\in\! \mathbb{R} }$,
satisfy the symmetries
(i) ${ U_{-k} (J,\Jp) \!=\! U_{k} (J,\Jp) }$;
(ii) ${ U_{k} (\Jp , J) \!=\! U_{k} (J , \Jp) }$.

We describe the system with its \DF\@,
${ F \!=\! F(\bw) }$, normalised to
${ \!\int\! \rd \bw F \!=\! \Mtot }$.
The system is in a quasi-stationary equilibrium,
so that both the mean \DF\@,
${ F \!=\! F (J) }$,
and the mean Hamiltonian,
${ H_{0} (\bw) \!=\! U_{\ext} (\bw) \!+\! \!\int\! \rd \bwp \, U (\bw,\bwp) F(\bwp) \!=\! H_{0} (J) }$,
depend only on the action.
The present system is said to be inhomogeneous because
(i) the coupling coefficients, ${ U_{k} (J , \Jp) }$,
depend on the particles' actions;
(ii) to every action is associated an orbital frequency, ${ \Omega (J) \!=\! \rd H_{0} / \rd J }$.
Characterising relaxation
amounts to characterising ${ \p F(J,t)/\p t }$.

In the limit where only ${1/N}$ effects are accounted for,
the dynamics of ${ F (J ,t) }$ is described
by the inhomogeneous \BL\ equation~\citep{Heyvaerts2010,Chavanis2012}.
It reads
\begin{align}
\frac{\p F (J)}{\p t} = 2 \pi^{2} \mu \frac{\p }{\p J} \bigg[ \!\! \int \!\! {} & \rd J_{1} \, \big| \psid_{\tot} (J , J_{1}) \big|^{2} \, \deltaD \big( \Omega[J] \!-\! \Omega [J_{1}] \big)
\nonumber
\\
\times {} & \bigg( \frac{\p }{\p J} \!-\! \frac{\p }{\p J_{1}} \bigg) \, F (J) \, F (J_{1}) \bigg] ,
\label{BL_eq}
\end{align}
where we dropped the time dependence of the \DFs\
for clarity.
Here, the total coupling coefficients follow from
${ |\psid_{\tot} (J, J_{1})|^{2} \!=\! \sum_{k} |k| | \psid_{kk}(J, J_{1}, k\Omega[J]) |^{2} }$,
where the detailed expression of the dressed coupling coefficients,
${ \psid_{k\kp} (J, J_{1}, \omega) }$, can be found in~\S{G}
of~\cite{Fouvry+2018}.
Importantly, we emphasise that the symmetry of Eq.~\eqref{def_Uk}
imposes ${ \psid_{k\kp} \!\propto\! \delta_{k\kp} }$,
i.e.\ only {1:1} resonances are permitted.

For a system with a monotonic frequency profile,
${ J \!\mapsto\! \Omega (J) }$, the diffusion flux
from Eq.~\eqref{BL_eq} exactly vanishes.
Indeed, the resonance condition, ${ \deltaD (\Omega[J] \!-\! \Omega[J_{1}]) }$,
imposes that only local two-body resonances of the form
${ J_{1} \!=\! J }$ are permitted.
This leads to ${ \p F/\p t \!=\! 0 }$ in Eq.~\eqref{BL_eq}.
As a consequence, one-dimensional inhomogeneous systems
with a monotonic frequency profile
and sustaining only {1:1} resonances
generically have a vanishing \BL\ flux.
This is a kinetic blocking~\citep[see, e.g.\@,][]{Barre+2014,Lourenco+2015},
i.e.\ these systems cannot relax
via two-body correlations of order ${1/N}$.
Relaxation is then significantly delayed
as it can only occur through the weaker ${1/N^{2}}$
three-body correlations.
This is our focus here.

In order to derive an appropriate ${1/N^2}$ kinetic equation
for the present system, we generalise the result from~\citep{Fouvry+2020},
which focused on homogeneous systems
with an arbitrary interaction potential.
Yet, accounting for inhomogeneity
makes it so that calculations become rapidly difficult to handle
given the large number of terms appearing.
All these aspects are dealt with
within a \texttt{Mathematica} code that is distributed
in the Supplemental Material~\cite{MMA}.

Building upon~\cite{Fouvry+2020},
the key steps of the derivation
are highlighted in~\S\ref{app:Derivation}.
In short, we proceed by:
(i) writing the usual BBGKY coupled evolution equations
for the one-, two- and three-body \DFs\@~\citep{Swanson2008};
(ii) rewriting these equations as evolution equations
for the one-body \DF\@, ${ F (J,t) }$,
and the two- and three-body correlation functions
using the cluster expansion~\citep{Balescu1997};
(iii) truncating these equations at order ${1/N^2}$
and splitting the two-body correlation function
in its ${1/N}$ and ${1/N^{2}}$ components~\citep{RochaFilho+2014};
(iv) neglecting collective effects
by assuming that the system is dynamically hot enough
so as to not strongly amplify its own self-generated
perturbations\footnote{It is the same assumption that allows one
to derive the (simpler) Landau equation from the \BL\ one, at order ${1/N}$~\citep{Chavanis2013}.};
(v) solving explicitly
a sequence of four differential equations.
Once these steps implemented,
it remains to perform a large number of
integration by parts, symmetrisations,
and relabellings
to reach a simple expression
for the final ${1/N^{2}}$ collision operator.
This is, by far, the most challenging part of the calculation
where the use of a computer algebra system
appears mandatory.
All the details are given in~\cite{MMA}.

Ultimately, the kinetic equation reads
\begin{align}
\frac{\p F (J)}{\p t} = {} & 2 \pi^{3} \mu^{2} \frac{\p }{\p J} \bigg[ \sum_{k_{1} , k_{2}} \!\! \frac{1}{k_{1}^{2} (k_{1} \!+\! k_{2})} \mP \!\!\! \int \!\! \frac{\rd J_{1}}{(\Omega[J] \!-\! \Omega[J_{1}])^{4}}
\nonumber
\\
\times {} & \!\! \int \!\! \rd J_{2} \, \mU_{k_{1}k_{2}} (\bJ) \, \deltaD \big[ \bk \!\cdot\! \bO \big] \, \bigg( \bk \!\cdot\! \frac{\p }{\p \bJ} \bigg) \, F_{3} (\bJ) \bigg] ,
\label{kin_eq}
\end{align}
where the sum over ${ k_{1} , k_{2} }$ is restricted to the integers such
that $k_{1}$, $k_{2}$ and ${ (k_{1} \!+\! k_{2}) }$ are all non-zero.
In that expression, we also shortened the notations
by introducing the vectors ${ \bJ \!=\! (J , J_{1} , J_{2})}$,
${ \bO \!=\! (\Omega[J],\Omega[J_{1}],\Omega[J_{2}]) }$,
${ \bk \!=\! (k_{1} \!+\! k_{2} , -k_{1} , - k_{2}) }$
as well as ${ F_{3} (\bJ) \!=\! F (J) F(J_{1}) F(J_{2}) }$.
Equation~\eqref{kin_eq} finally involves the (positive)
coupling coefficients, ${ \mU_{k_{1}k_{2}} (\bJ) }$,
whose detailed expression is given in~\S\ref{app:mU}.
We note that Eq.~\eqref{kin_eq} is sourced
by three-body correlations,
i.e.\ it involves the \DF\ three times.
Therein, relaxation occurs only when the three-body resonance condition,
${ \deltaD [\bk \!\cdot\! \bO ] }$, is satisfied.
Finally, Eq.~\eqref{kin_eq} also differs from Eq.~\eqref{BL_eq}
because it does not involve collective effects.

In the limit of an homogeneous system,
one may replace
(i) the action $J$ by the velocity $v$;
(ii) the orbital frequency, ${ \Omega (J) }$, by $v$;
(iii) the action-dependent coefficients, ${ U_{k} (J,\Jp) }$,
by the velocity-independent ones, $U_{k}$.
Following such replacements,
one exactly recovers the homogeneous ${1/N^{2}}$
kinetic equation already derived in~\cite{Fouvry+2020}.

Equation~\eqref{kin_eq} is the main result of this work.
It describes the (very) long-term relaxation
of dynamically hot one-dimensional inhomogeneous systems,
as driven by ${1/N^{2}}$ effects.
Equation~\eqref{kin_eq} is rather general
since it applies to any arbitrary long-range interaction potential
that follows Eq.~\eqref{def_Uk}. In the coming section, we explore
some of the main properties of this kinetic equation.

\section{Properties}
\label{sec:Properties}

\subsection{Conservation laws}
\label{sec:Conservation}

Equation~\eqref{kin_eq} satisfies a couple
of important conservation laws. Up to prefactors,
those are
\begin{align}
M (t) {} & = \!\! \int \!\! \rd J \, F (J , t) 
{} & \text{(total mass)} ;
\nonumber
\\
P (t) {} & = \!\! \int \!\! \rd J \, J \, F (J , t) 
{} & \text{(total momentum)} ;
\nonumber
\\
E (t) {} & = \!\! \int \!\! \rd J  \, H_{0} (J) \, F (J , t) 
{} & \text{(total energy)} .
\label{conservation}
\end{align}
We refer to~\S\ref{app:Cons} for the proof
of these conservations.

\subsection{Dimensionless rescaling}
\label{sec:Dimensionless}

It is enlightening to estimate the typical relaxation time
predicted by Eq.~\eqref{kin_eq} using a dimensionless rewriting.
We denote the system's typical frequency with $\Omega_{0}$
and set the dynamical time to ${ \Td \!=\! 1 / \Omega_{0} }$.
We define the typical action $J_{0}$
via the \DF\@'s action dispersion,
and introduce the dimensionless \DF\ ${ \oF \!=\! J_{0} F / \Mtot }$.
Finally, we assume that the interaction potential
satisfies ${ U \!\propto\! G }$.
Injecting these various elements in Eq.~\eqref{kin_eq},
we find ${ \p \oF / \p \ot \!=\! 1/(N^{2} Q^{4}) ... }$
with ${ \ot \!=\! t/\Td }$ and all the remaining terms dimensionless.
Here, we introduced the system's dimensionless stability parameter~\citep[see, e.g.\@,][]{Toomre1964}
\begin{equation}
Q = \frac{J_{0} \, \Omega_{0}}{G \, \Mtot} .
\label{def_Q}
\end{equation}
The larger $Q$, the hotter the system,
i.e.\ the weaker the collective effects\footnote{The typical
scaling of $Q$ can readily be found
from the system's inhomogeneous response matrix~\citep[see~\S{5.3} in][]{BinneyTremaine2008}.}.
The system's relaxation time, $\Tr$,
when driven by Eq.~\eqref{kin_eq}, therefore scales like
\begin{equation}
\Tr \simeq Q^{4} N^{2} \Td .
\label{def_Tr}
\end{equation}
As expected, the hotter the system,
the slower the long-term relaxation.
Given that Eq.~\eqref{kin_eq} has been derived
by neglecting collective effects, it can only
be applied to systems with ${ Q \!\gg\! 1 }$.

\subsection{Well-posedness}
\label{sec:WellPosed}

Because it involves a high-order resonance denominator,
it is not obvious that Eq.~\eqref{kin_eq}
is well-posed, i.e.\ that there are no divergences
when ${ J_{1} \!\to\! J }$. Following the same approach
as~\cite{Fouvry+2020},
we show in~\S\ref{app:WellPosed}
that one can rewrite Eq.~\eqref{kin_eq}
under an alternative form for which
the principal value can be computed.

\section{Steady states}
\label{sec:Steady}

\subsection{$H$-theorem}
\label{sec:HTheorem}

Up to prefactors, the system's entropy is defined as
\begin{equation}
S (t) = - \!\! \int \!\! \rd J \, s \big[ F (J , t) \big] ,
\label{def_S}
\end{equation}
with ${ s [F] \!=\! F \ln (F) }$ Boltzmann's entropy.
As detailed in~\S\ref{app:HTheorem},
one can show that Eq.~\eqref{kin_eq} drives an evolution
of the entropy according to
\begin{align}
\frac{\rd S}{\rd t} {} & \!=\! \frac{2 \pi^{3} \mu^{2}}{3} \!\!\sum_{\mathclap{k_{1} , k_{2}}} \! \int \!\! \rd \bJ \frac{\mU_{k_{1} k_{2}} (\bJ)}{k_{1}^{2} (k_{1} \!+\! k_{2})^{2}} \mP \bigg(\!\! \frac{1}{(\Omega[J] \!-\! \Omega [J_{1}])^{4}} \!\!\bigg)
\label{dSdt}
\\
{} & \times \frac{\deltaD [\bk \!\cdot\! \bO]}{F_{3} (\bJ)} \bigg(\!\! (k_{1} \!+\! k_{2}) \frac{F' \!(J)}{F (J)} \!-\! k_{1} \frac{F' \!(J_{1})}{F (J_{1})} \!-\! k_{2} \frac{F' \!(J_{2})}{F (J_{2})} \!\bigg)^{2} .
\nonumber
\end{align}
All the terms in this integral are positive.
Hence, Eq.~\eqref{kin_eq} satisfies an $H$-theorem,
i.e.\
\begin{equation}
\frac{\rd S}{\rd t} \geq 0 .
\label{dSdt_pos}
\end{equation}
Equation~\eqref{kin_eq} therefore drives an irreversible relaxation.

\subsection{Boltzmann distribution}
\label{sec:BoltzmannDF}

For the present case, the thermodynamical equilibria,
i.e.\ the Boltzmann \DFs\@,
are generically of the form
\begin{equation}
\FB (J) \propto \re^{- \beta H_{0} (J) + \gamma J} ,
\label{generic_Boltzmann}
\end{equation}
with ${ \beta , \gamma }$
two Lagrange multipliers
associated with the conservation of ${ E(t) }$ and ${ P (t) }$
in Eq.~\eqref{conservation}.
When injected in Eq.~\eqref{kin_eq},
the \DF\ from Eq.~\eqref{generic_Boltzmann} gives
\begin{align}
\frac{\p \FB (J)}{\p t} {} & \!\propto\! \deltaD [\bk \!\cdot\! \bO] \,
\big\{ \!-\! \beta \, \big[ \bk \!\cdot\! \bO \big] \!+\! \gamma \, \big[ (k_{1} \!+\! k_{2}) \!-\! k_{1} \!-\! k_{2} \big] \big\}
\nonumber
\\
{} & = 0 .
\label{rate_dFBdt}
\end{align}
As expected, Boltzmann \DFs\ are
equilibria of Eq.~\eqref{kin_eq}.

\subsection{Constraint from the $H$-theorem}
\label{sec:HtheoremCons}

Following the calculation of ${ \rd S / \rd t }$ in Eq.~\eqref{dSdt},
let us now investigate what are the most generic steady states
of Eq.~\eqref{kin_eq}. For simplicity, we assume that there exists
${ (k_{1},k_{2}) }$ such that ${ \mU_{k_{1} k_{2}} (\bJ) \!\neq\! 0 }$
when ${ \bk \!\cdot\! \bO \!=\! 0 }$, i.e.\ at resonance.
An obvious way of ensuring ${ \rd S / \rd t \!=\! 0 }$
is for the last term in Eq.~\eqref{dSdt} to systematically vanish.
Since ${ J \!\mapsto\! \Omega (J) }$ is monotonic,
we can define the function ${ G (\Omega) \!=\! F'(J [\Omega])/F(J[\Omega]) }$
and we find the constraint
\begin{equation}
\forall \Omega_{1},\Omega_{2} \!:\,
G \bigg(\! \frac{k_{1} \Omega_{1} \!+\! k_{2} \Omega_{2}}{k_{1} \!+\! k_{2}} \!\bigg) = \frac{k_{1} G (\Omega_{1}) \!+\! k_{2} G (\Omega_{2})}{k_{1} \!+\! k_{2}} ,
\label{constraint_G}
\end{equation}
namely a weighted average. For this constraint to be satisfied
for all $\Omega_{1}$, $\Omega_{2}$, the function
${ \Omega \!\mapsto\! G (\Omega) }$ must necessarily be affine,
i.e.\ one has
\begin{equation}
G (\Omega) = - \beta \, \Omega + \gamma .
\label{line_G}
\end{equation}
Integrating once w.r.t.\ $J$, Eq.~\eqref{line_G}
recovers the Boltzmann \DF\ from Eq.~\eqref{generic_Boltzmann}.
Provided that there exists one ${ (k_{1} , k_{2}) }$
for which ${ \mU_{k_{1} k_{2}} (\bJ) \!\neq\! 0 }$ at resonance,
the only equilibrium \DFs\ of Eq.~\eqref{kin_eq}
are the Boltzmann \DFs\@.

\subsection{Second-order kinetic blocking}
\label{sec:SecondKB}

It is interesting to determine whether
or not one can design an interaction potential
and a frequency profile so that, for all ${ (k_{1} , k_{2}) }$
one has ${ \mU_{k_{1}k_{2}} (\bJ) \!=\! 0 }$
at resonance.
Following Eq.~\eqref{dSdt},
this would then impose ${ \rd S / \rd t \!=\! 0 }$
whatever the considered \DF\@.
For simplicity, in this section,
we assume the simple frequency profile
${ \Omega (J) \!\propto\! J }$.

In~\S\ref{app:2nd}, 
we show that potentials of the form
\begin{equation}
U (\bw , \bwp) \propto \big| J \!-\! \Jp \big|^{\alpha} \sum_{\mathclap{\substack{k = 1 \\ k \equiv 0 \, [\mathrm{mod} \, d]}}}^{+ \infty} \frac{1}{|k|^{\alpha}} \, \cos \!\big[ k \, (\theta \!-\! \thetap) \big] ,
\label{generic_U_w}
\end{equation}
with $\alpha$ an arbitrary power law index
and ${ d \!\geq\! 1 }$ an arbitrary integer,
are generically blocked
for the ${ 1/N^{2} }$ dynamics driven by Eq.~\eqref{kin_eq}.
More precisely, such potentials ensure that the coupling function
${ \mU_{k_{1}k_{2}} (\bJ) }$ (\S\ref{app:mU}) satisfies
\begin{equation}
\forall k_{1}, k_{2}, J, J_{1} \!\!:\,
\mU_{k_{1}k_{2}} (J,J_{1},J_{2}^{\res}) = 0 ,
\label{vanish_U}
\end{equation}
with ${ J_{2}^{\res} \!=\!  J_{2}^{\res} [\bk , J , J_{1}]}$,
the resonant action
complying with the constraint ${ \bk \!\cdot\! \bO \!=\! 0 }$.
For the frequency profile ${ \Omega [J] \!\propto\! J }$,
it simply reads
${ J_{2}^{\res} \!=\! [(k_{1} \!+\! k_{2}) J \!-\! k_{1} J_{1} ]/ k_{2}}$.
In~\S\ref{app:2nd}, we show that (i)
Eq.~\eqref{generic_U_w} allows one to recover
the local interaction potentials already unveiled in~\cite{Fouvry+2020};
(ii) for particular values of $\alpha$, the harmonic sum from
Eq.~\eqref{generic_U_w} can be explicitly performed.

Following Eq.~\eqref{generic_U_w},
\textit{any} (stable) \DF\@,
${ F \!=\! F (J) }$, when embedded within the mean potential
${ H_{0} \!=\! \half J^{2}  }$ and pairwise interaction
from Eq.~\eqref{generic_U_w},
satisfies ${ \p F / \p t \!=\! 0 }$
when plugged into Eq.~\eqref{kin_eq}.
We call such a situation a second-order kinetic blocking,
i.e.\ both the ${1/N}$ \BL\ flux
and the ${ 1/N^2 }$ flux from Eq.~\eqref{kin_eq}
exactly vanish.
In the limit where collective effects can effectively be neglected,
we expect that such systems will relax
on the much longer timescales
${ \mO (N^{3} \Td) }$.
Given the difficulty of (i) deriving a kinetic equation at order ${1/N^3}$;
and (ii) numerically integrating such dynamics
on extremely long timescales,
we postpone their investigation to future works.

\section{Application}
\label{sec:Application}

In order to test Eq.~\eqref{kin_eq},
we perform numerical simulations
of classical Heisenberg spins
embedded within an external potential~\citep[see, e.g.\@,][]{Gupta+2011,Barre+2014}.
We refer to~\S\ref{app:NBody}
for a detailed presentation of the setup.

\begin{figure}[htbp!]
    \begin{center}
    \includegraphics[width=0.45\textwidth]{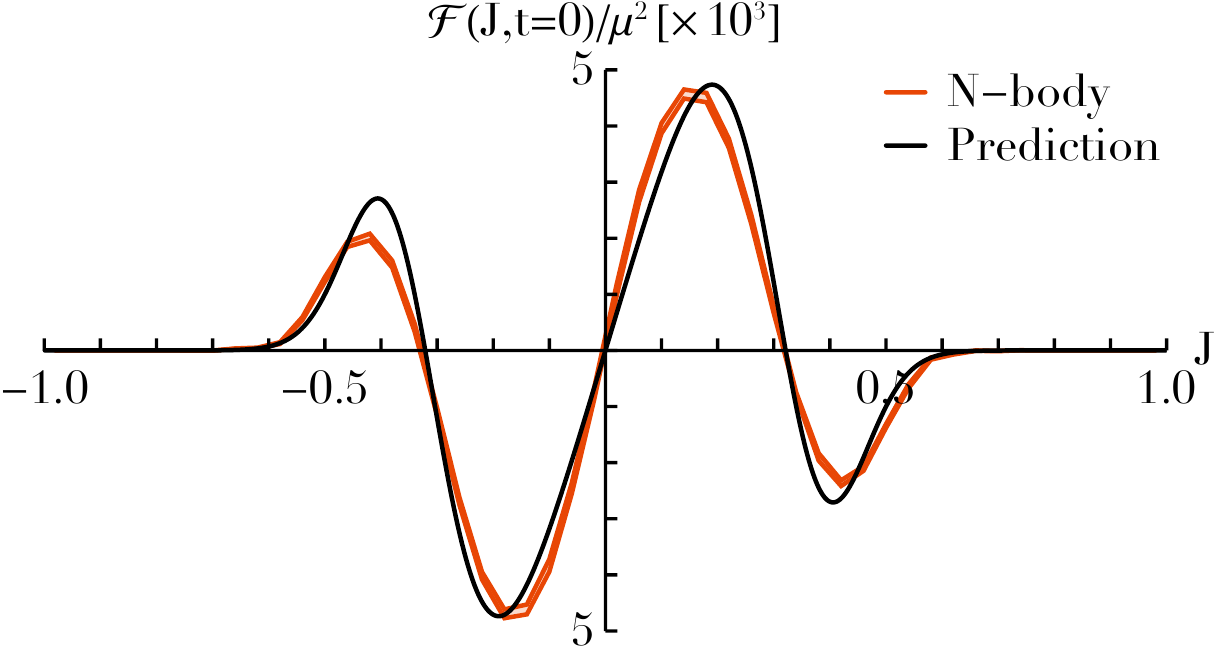}
    \caption{Initial diffusion flux, ${ \p F / \p t \!=\! \p \mF / \p J }$,
    in action space
    as measured in direct $N$-body simulations
    and predicted by Eq.~\eqref{kin_eq}.
    Detailed parameters are given in~\S\ref{app:NBody}.
    \label{fig:Flux}}
    \end{center}
\end{figure}

In Fig.~\ref{fig:Flux}, we illustrate the initial diffusion flux,
${ \p F / \p t \!=\! \p \mF / \p J }$.
This figure shows a good quantitative agreement
between the predicted and measured fluxes.
The remaining slight mismatch likely stems from
(i) non-vanishing contributions from the source term
in ${ G_{2}^{(1)} \!\times\! G_{2}^{(1)} }$, see~\S\ref{app:TruncN2};
(ii) remaining contributions associated with collective effects.

\section{Conclusions}
\label{sec:Conclusions}

We presented a ${ 1/N^{2} }$
closed and explicit kinetic equation
for long-range interacting one-dimensional inhomogeneous systems.
The collision operator from Eq.~\eqref{kin_eq}
generalises the classical
inhomogeneous Landau kinetic equation
to regimes where the ${ 1/N }$ relaxation exactly vanishes
by symmetry.
Equation~\eqref{kin_eq} conserves the total mass, momentum
and energy, and satisfies an $H$-theorem.
We exhibited a class of long-range interaction potentials
for which this ${1/N^2}$ collision term exactly vanishes.
Finally, we showed how Eq.~\eqref{kin_eq} quantitatively
matches with measurements from direct $N$-body simulations.

Naturally, the present work is only one step toward
ever more detailed characterisations of the long-term relaxation
of finite-$N$ systems. As such it would be worthwhile to
(i) generalise Eq.~\eqref{kin_eq} to also account for collective effects
--  a significantly challenging endeavour;
(ii) account for the source term in
${ G_{2}^{(1)} \!\times\! G_{2}^{(1)} }$~(\S\ref{app:TruncN2});
(iii) investigate, both analytically and numerically,
systems driven by the interaction potential
from Eq.~\eqref{generic_U_w},
for which we expect a second-order kinetic blocking,
i.e.\ a relaxation on ${ N^{3} \Td }$ timescales.

\begin{acknowledgments}
This work is partially supported by the grant Segal ANR-19-CE31-0017 
of the French Agence Nationale de la Recherche
and by the Idex Sorbonne Universit\'e.
We thank S.\ Rouberol for the smooth running
of the Infinity cluster,
where the simulations were performed.
We thank M.\ Roule, M.\ Petersen and C.\ Pichon
for many stimulating discussions.
\end{acknowledgments}

\appendix

\section{Derivation}
\label{app:Derivation}

Most of the derivation of Eq.~\eqref{kin_eq}
follows the same lines
as in~\cite{Fouvry+2020} (see~\S{A} therein).
Here, we only highlight the key changes
stemming from inhomogeneity.

\vspace{-6mm}
\subsection{BBGKY hierarchy}
\label{app:BBGKY}
\vspace{-2mm}

The dynamics of the system is exactly
described by the BBGKY equations for the $n$-body \DFs\@,
${F_{n} (\bw_{1},...,\bw_{n},t)}$~\cite[see~\S{A1} in][]{Fouvry+2020}.
It reads
\begin{equation}
\frac{\p F_{n}}{\p t} + \big[ F_{n} , H_{n} \big]_{n} + \!\! \int \!\! \rd \bw_{n+1} \, \big[ F_{n+1} , \delta H_{n+1} \big]_{n} = 0 ,
\label{BBGKY_n}
\end{equation}
with the Poisson bracket
\begin{equation}
\big[ f , h \big]_{n} = \sum_{i=1}^{n} \bigg( \frac{\p f}{\p \theta_{i}} \frac{\p h}{\p J_{i}} - \frac{\p f}{\p J_{i}} \frac{\p h}{\p \theta_{i}} \bigg) .
\label{def_Poisson_bracket}
\end{equation}
In Eq.~\eqref{BBGKY_n}, the $n$-body Hamiltonian,
$H_{n}$, follows from Eq.~\eqref{Htot_init}
with the replacement ${ N \!\to\! n }$.
In Eq.~\eqref{BBGKY_n}, the specific interaction energy,
${ \delta H_{n+1} }$, simply reads
\begin{equation}
\delta H_{n+1} (\bw_{1} , ... , \bw_{n+1}) = \sum_{i = 1}^{N} U (\bw_{i} , \bw_{n+1}) .
\label{def_deltaH}
\end{equation}
Importantly, Eq.~\eqref{BBGKY_n} provides us with
the evolution equations for the one-, two-, and three-body
\DFs\ ($F_{1}$, $F_{2}$ and $F_{3}$).
This is the starting point of the derivation.

\subsection{Cluster expansion}
\label{app:Cluster}

To perform perturbative expansions
w.r.t.\ the small parameter ${1/N}$, the $n$-body
\DFs\ are developed using the cluster expansion~\citep{Balescu1997}.
Doing so, one introduces the two- and three-body correlation
functions, ${ G_{2} (\bw_{1},\bw_{2}) }$ and ${ G_{3} (\bw_{1} , \bw_{2} , \bw_{3}) }$.
As an example, $G_{2}$ is defined via
\begin{equation}
F_{2} (\bw_{1} , \bw_{2}) = F (\bw_{1}) \, F (\bw_{2}) + G_{2} (\bw_{1} , \bw_{2}) ,
\label{def_G2}
\end{equation}
where, from now on, we write the one-body \DF\
as ${ F \!=\! F_{1} }$.
Once this expansion performed, the quantities at our disposal
scale w.r.t.\ $N$ like
${ F \!\sim\! 1 }$,
${ G_{2} \!\sim\! 1/N }$,
${ G_{3} \!\sim\! 1/N^{2} }$.

These expansions are then injected into Eq.~\eqref{BBGKY_n}
to obtain evolution equations for ${ \p F / \p t }$,
${ \p G_{2} / \p t }$ and ${ \p G_{3} / \p t }$.
All these calculations are explicitly performed in~\cite{MMA}.

\subsection{Truncation at order ${ 1/N^{2} }$}
\label{app:TruncN2}

The next step
is to truncate these three evolution equations
at order ${ 1/N^{2} }$.
We perform the operations:

(I) We introduce the small parameter ${ \eps \!=\! 1/N }$
and decompose the two-body correlation function as
\begin{equation}
G_{2} = \eps \, G_{2}^{(1)} + \eps^{2} \, G_{2}^{(2)} .
\label{decomposition_G2}
\end{equation}
Similarly, the other parameters at our disposal
are rescaled as ${ \mu \!\to\! \eps \, \mu }$,
${ G_{3} \!\to\! \eps^{2} \, G_{3} }$.

(II) We truncate the BBGKY evolution equations
up to order $\eps^{2}$.
We split the evolution equation for ${ \p G_{2} / \p t }$
into two, respectively for ${ \p G_{2}^{(1)}/ \p t }$
and ${ \p G_{2}^{(2)} / \p t }$.

(III) We leverage our assumption of a quasi-stationarity,
i.e.\ ${ F \!=\! F (J) }$ and ${ H_{0} \!=\! H_{0} (J) }$,
hence introducing the orbital frequencies, ${ \Omega (J) }$.

(IV) We neglect the contributions from collective effects,
i.e.\ we perform replacements of the form
\begin{equation}
\!\! \int \!\! \rd \bw_{3} \, G_{2}^{(1)} (\bw_{2} , \bw_{3}) \, \p_{\theta_{1}} U(\bw_{1} , \bw_{3}) \to 0 ,
\label{neglect_CollEff}
\end{equation}
and similarly for $G_{2}^{(2)}$ and $G_{3}$.

(V) In the hot limit, we neglect the source term
in ${ G_{2}^{(1)} \!\times\! G_{2}^{(1)} }$
in ${ \p G_{3} / \p t }$,
as its contribution is a factor ${1/Q}$ smaller
than the source term in $G_{2}^{(2)}$.

\subsection{Solving the equations}
\label{app:SolvingEq}

Following all these steps, we have at our disposal
a set of four coupled partial differential equations.
These equations can be solved sequentially,
starting with $G_{2}^{(1)}$, then $G_{3}$, $G_{2}^{(2)}$
and finally $F$.
This is done as follows:

(I) We perform Fourier expansions w.r.t.\ the angles $\theta_{i}$.

(II) We rely on the assumption of timescale separation,
hence fixing ${ F (J,t) \!=\! \cst }$ when solving
for fluctuations.

(III) We impose the initial conditions
${ G_{2}^{(1)} (t \!=\! 0) \!=\! 0 }$
and similarly for $G_{2}^{(2)}$ and $G_{3}$.

Ultimately, we obtain an explicit time-dependent expression
for ${ G_{2}^{(2)} (t) }$.
Leveraging once again timescale separation,
we consider the limit ${ t \!\to\! + \infty }$
of ${ G_{2}^{(2)} (t) }$ in the evolution equation
for ${ \p F / \p t }$.
A typical time integral is then replaced asymptotically by
\begin{equation}
\lim\limits_{t \to + \infty} \!\! \int_{0}^{t} \!\! \rd t_{1} \, \re^{\ri (t - t_{1}) \omegaR} = \pi \deltaD (\omegaR) + \ri \, \mP \bigg( \frac{1}{\omegaR} \bigg) ,
\label{Plemelj_formula}
\end{equation}
where ${ \omegaR \!\in\! \mathbb{R} }$
is a linear combination of ${ \bO }$.

\subsection{Simplifying the expressions}
\label{app:Simplifications}

At this stage, we are left with a kinetic equation involving thousands of terms.
The computer algebra system allows
for efficient manipulations
of these expressions.
The key steps are as follows:

(I) We implement systematic relabellings of the actions ${ (J_{1},J_{2}) }$
and resonance numbers ${ (k_{1},k_{2}) }$, so that the resonant frequencies
are all of the form ${ \omegaR \!=\! \bk \!\cdot\! \bO }$.

(II) We integrate by parts w.r.t.\ the actions so that
no derivatives act on $\deltaD$,
and only first-order derivatives of the \DF\
and coupling coefficients are present.

(III) We use the scaling relations of $\deltaD$ and $\mP$,
e.g.\@, ${ \deltaD (\alpha x) \!=\! \deltaD(x)/|\alpha| }$,
to take out resonance numbers.

(IV) The frequency profile being monotonic,
we use
\begin{equation}
\!\! \int \!\! \rd J_{2} \, f (J_{1} , J_{2}) \, \deltaD \big( \Omega [J_{1}] \!-\! \Omega[J_{2}] \big) = \frac{f(J_{1} , J_{1})}{|\Omega' (J_{1})|} .
\label{replacement_Dirac}
\end{equation}

(V) We use the resonance condition
${ \deltaD [\bk \!\cdot\! \bO] }$ to make the replacements
${ (\Omega \!-\! \Omega_{2}) \!\to\! - (k_{1}/k_{2}) (\Omega \!-\! \Omega_{1}) }$
and ${ (\Omega_{1} \!-\! \Omega_{2}) \!\to\! - ([k_{1} \!+\! k_{2}]/k_{2}) (\Omega \!-\! \Omega_{1})}$,
with the shortened notation ${ (\Omega , \Omega_{1} , \Omega_{2}) \!=\! \bO }$.
Principal values are therefore only expressed
as functions of ${ (\Omega \!-\! \Omega_{1}) }$.

All these manipulations are automated
using tailored rules in \texttt{Mathematica},
that can all be found in~\cite{MMA}.
Ultimately, one obtains the closed result from Eq.~\eqref{kin_eq}.

\section{Coupling coefficients}
\label{app:mU}

The coupling coefficients, ${ \mU_{k_{1}k_{2}} (\bJ) }$,
appearing in Eq.~\eqref{kin_eq} are generically given by
\begin{equation}
\mU_{k_{1}k_{2}} (\bJ) \!=\! \big[ (\Omega[J] \!-\! \Omega[J_{1}]) \, \mU_{k_{1}k_{2}}^{(1)} (\bJ) + k_{2} \, \mU^{(2)}_{k_{1}k_{2}} (\bJ) \big]^{2} .
\label{exp_mU}
\end{equation}
The coupling functions appearing in this expression read
\begin{align}
\mU_{k_{1} k_{2}}^{(1)} (\bJ) =
{} & k_{2} (k_{1} \!+\! k_{2}) \big\{ U_{k_{1} + k_{2}} \!(J , J_{2}) \, \p_{J_{2}} U_{k_{1}} \!(J_{1} , J_{2})
\nonumber
\\
{} & \;\;\;\;\;\;\;\;\; - U_{k_{2}} \!(J , J_{2}) \, \p_{J} U_{k_{1}} \!(J , J_{1}) \big\}
\nonumber
\\
+ {} & k_{1} (k_{1} \!+\! k_{2}) \big\{ U_{k_{1}} \!(J , J_{1}) \, \p_{J} U_{k_{2}} \!(J , J_{2})
\nonumber
\\
{} & \;\;\;\;\;\;\;\;\; - U_{k_{1} + k_{2}} \!(J , J_{1}) \, \p_{J_{1}} U_{k_{2}} \!(J_{1} , J_{2}) \big\}
\nonumber
\\
- {} & k_{1} k_{2} \big\{ U_{k_{2}} \!(J_{1} , J_{2}) \, \p_{J_{1}} U_{k_{1} + k_{2}} \!(J , J_{1}) 
\nonumber
\\
{} & \;\;\;\;\;\;\;\;\; - U_{k_{1}} \!(J_{1} , J_{2}) \, \p_{J_{2}} U_{k_{1} + k_{2}} \!(J , J_{2}) \big\} ,
\label{def_U1}
\end{align}
and
\begin{align}
\mU_{k_{1} k_{2}}^{(2)} (\bJ) \!=\! (k_{1} \!+\! k_{2}) \, {} & \frac{\rd \Omega}{\rd J} \, U_{k_{1}} (J , J_{1}) \, U_{k_{2}} (J , J_{2}) 
\nonumber
\\
- k_{1} \, {} & \frac{\rd \Omega}{\rd J_{1}} \, U_{k_{1} + k_{2}} (J , J_{1}) \, U_{k_{2}} (J_{1} , J_{2}) 
\nonumber
\\
- k_{2} \, {} & \frac{\rd \Omega}{\rd J_{2}} \, U_{k_{1}} (J_{1} , J_{2}) \, U_{k_{1} + k_{2}} (J , J_{2}) .
\label{def_U2}
\end{align}
We refer to~\cite{MMA} for the associated derivation.

\section{Properties}
\label{app:Properties}

\subsection{Conservation laws}
\label{app:Cons}

We can generically rewrite Eq.~\eqref{kin_eq} as
\begin{equation}
\frac{\p F (J)}{\p t} = \frac{\p \mF (J)}{\p J} ,
\label{def_Flux}
\end{equation}
with ${ \mF (J) }$ the diffusion flux.
The time derivatives of Eq.~\eqref{conservation}
then read
\begin{align}
\frac{\rd M}{\rd t} {} & = \!\! \int \!\! \rd J \, \frac{\p \mF}{\p J} ,
\nonumber
\\
\frac{\rd P}{\rd t} {} & = - \!\! \int \!\! \rd J \, \mF (J) ,
\nonumber
\\
\frac{\rd E}{\rd t} {} & = - \!\! \int \!\! \rd J \, \Omega(J) \, \mF (J) .
\label{calc_der_cons}
\end{align}
The conservation of the total mass, ${ M (t) }$,
follows from the absence of any boundary contributions.

For the conservation of ${ P (t) }$ and ${ E(t) }$,
we investigate
\begin{equation}
\!\! \int \!\! \rd J \, \mF (J) = \sum_{\mathclap{k_{1} , k_{2}}} (k_{1} \!+\! k_{2}) \, \!\! \int \!\! \rd \bJ \, A_{k_{1}k_{2}} (\bJ) ,
\label{sym_Flux_1}
\end{equation}
with ${ A_{k_{1}k_{2}} (\bJ) }$
given by Eq.~\eqref{kin_eq}.
With the relabellings ${ J \!\leftrightarrow\! J_{1} }$
and ${ (k_{1} ,k_{2}) \!\to\! (-k_{1} \!-\! k_{2},k_{2}) }$,
Eq.~\eqref{sym_Flux_1} becomes~\citep{MMA}
\begin{equation}
\!\! \int \!\! \rd J \, \mF (J) = - \sum_{\mathclap{k_{1},k_{2}}} k_{1} \!\! \int \!\! \rd \bJ \, A_{k_{1}k_{2}} (\bJ) .
\label{sym_Flux_2}
\end{equation}
Similarly, with the relabellings ${ J \!\leftrightarrow\! J_{2} }$
and ${ (k_{1},k_{2}) \!\to\! (-k_{1},k_{1}\!+\!k_{2}) }$,
Eq.~\eqref{sym_Flux_1} becomes~\citep{MMA}
\begin{equation}
\!\! \int \!\! \rd J \, \mF (\bJ) = - \sum_{\mathclap{k_{1} , k_{2}}} k_{2} \!\! \int \!\! \rd \bJ \, A_{k_{1}k_{2}} (\bJ) .
\label{sym_Flux_3}
\end{equation}

We can now go back to the computation
of the conserved quantities in Eq.~\eqref{calc_der_cons}.
By adding ${ \third }$ of Eqs.~\eqref{sym_Flux_1}--\eqref{sym_Flux_3},
we finally obtain
\begin{align}
\frac{\rd P}{\rd t} {} & = - \third \sum_{\mathclap{k_{1},k_{2}}} \!\! \int \!\! \rd \bJ \, A_{k_{1}k_{2}}(\bJ) \, \big\{ (k_{1} \!+\! k_{2}) \!-\! k_{1} \!-\! k_{2} \big\} = 0,
\nonumber
\\
\frac{\rd E}{\rd t} {} & = - \third \sum_{\mathclap{k_{1} , k_{2}}} \!\! \int \!\! \rd \bJ \, A_{k_{1} k_{2}} (\bJ) \, \big\{ \bk \!\cdot\! \bO \big\} = 0 ,
\label{final_calc_cons}
\end{align}
where the final equality stems from the presence
of the resonance condition, ${ \deltaD [\bk \!\cdot\! \bO] }$,
in Eq.~\eqref{kin_eq}.

\subsection{Well-posedness}
\label{app:WellPosed}

Following the same approach as in~\citep{Fouvry+2020},
we define the set of fundamental resonances as
\begin{equation}
\big\{ (k , \kp) \,|\, 0 < k , \kp \big\} .
\label{fund_res}
\end{equation}
For a given fundamental resonance, ${ (k,\kp) }$,
we define the associated set of resonance pairs,
${ (k_{1},k_{2}) }$, with
\begin{align}
\mR (k,\kp) = \big\{ {} & (k,\kp) , (k \!+\! \kp , - k) , (k,-k \!-\! \kp) ,
\nonumber
\\
{} & (\kp,k) , (k\!+\!\kp,-\kp) , (\kp,-k\!-\!\kp) \big\} .
\label{def_mR}
\end{align}
We note that (i) all ${ (k_{1} , k_{2}) }$
in ${ \mR (k,\kp) }$ satisfy ${ k_{1} \!\geq\! 0 }$;
(ii) even for ${ k \!=\! \kp }$,
this set still contains six elements.

Following these definitions, we rewrite Eq.~\eqref{kin_eq} as~\citep{MMA}
\begin{equation}
\frac{\p F (J)}{\p t} = 2 \pi^{3} \mu^{2} \frac{\p }{\p J} \bigg[ \,\, \sum_{\mathclap{k,\kp > 0}} \mF_{k\kp}(J) \bigg] ,
\label{rewrite_kin_eq_Fund}
\end{equation}
where ${ \mF_{k\kp} (J) }$ stands for the flux generated
by the fundamental resonance ${ (k,\kp) }$. It reads
\begin{equation}
\mF_{k\kp} (J) = \mP \!\! \int \!\! \frac{\rd J_{1}}{(\Omega[J] \!-\! \Omega[J_{1}])^{4}} \, \sum_{\mathclap{\substack{\\[0.5ex](k_{1},k_{2}) \in \mR (k,\kp)}}} \mC_{k_{1}k_{2}} (J , J_{1}) .
\label{def_mF_Fund}
\end{equation}
Here, ${ \mC_{k_{1}k_{2}} (J,J_{1}) }$ follows
from Eq.~\eqref{kin_eq} and reads
\begin{equation}
\mC_{k_{1}k_{2}} (J,J_{1}) = \!\! \int \!\! \rd J_{2} \, \frac{\mU_{k_{1}k_{2}} (\bJ)}{k_{1}^{2} (k_{1} \!+\! k_{2})} \, \deltaD [\bk \!\cdot\! \bO] \, \bigg( \bk \!\cdot\! \frac{\p }{\p \bJ} \bigg) \,  F_{3} (\bJ) .
\label{def_mC}
\end{equation}

Combining Eqs.~\eqref{def_mF_Fund} and~\eqref{def_mC},
we must ultimately perform an integration w.r.t.\ ${ \rd J_{1} \rd J_{2} }$.
As in~\cite{Fouvry+2020},
we can propose an alternative writing for ${ \mC_{k_{1}k_{2}} (J,J_{1}) }$
by performing the relabelling ${ J_{1} \!\leftrightarrow\! J_{2} }$.
We obtain~\citep{MMA}
\begin{equation}
\omC_{k_{1}k_{2}} (J,J_{1}) = \!\! \int \!\! \rd J_{2} \, \frac{\omU_{k_{1}k_{2}} (\bJ)}{k_{2}^{2} (k_{1} \!+\! k_{2})} \, \deltaD [\obk \!\cdot\! \bO] \, \bigg( \obk \!\cdot\! \frac{\p }{\p \bJ} \bigg) \, F_{3} (\bJ) .
\label{def_omC}
\end{equation}
As in Eq.~\eqref{kin_eq}, we introduced
${ \obk \!=\! (k_{1} \!+\! k_{2} , - k_{2} , -k_{1}) }$
along with the coupling function
\begin{equation}
\omU_{k_{1}k_{2}} \!(\bJ) \!=\! \big[ (\Omega[J] \!-\! \Omega[J_{1}]) \, \omU_{k_{1}k_{2}}^{(1)} \!(\bJ) \!-\! k_{1} \omU^{(2)}_{k_{1}k_{2}} \!(\bJ) \big]^{4} .
\label{def_omU}
\end{equation}
In that expression, ${ \omU_{k_{1}k_{2}}^{(1)} (\bJ) }$
(resp.\ ${ \omU_{k_{1}k_{2}}^{(2)} (\bJ) }$)
are directly obtained from Eq.~\eqref{def_U1}
(resp.\ Eq.~\ref{def_U2}) by performing
the relabelling ${ J_{1} \!\leftrightarrow\! J_{2} }$.

To obtain a well-posed expression
for the flux, ${ \mF_{k\kp} (J) }$,
it remains to appropriately combine the two possible writings,
${ \mC_{k_{1}k_{2}} (J,J_{1}) }$ and ${\omC_{k_{1}k_{2}} (J,J_{1})}$.
Once again, the computer algebra system
proves undoubtedly useful~\citep{MMA}.
Starting from Eqs.~\eqref{def_mF_Fund},
we consider the writing
\begin{align}
\mF_{k\kp} (J) = {} & \mP \!\! \int \!\! \frac{\rd J_{1}}{(\Omega[J] \!-\! \Omega[J_{1}])^{4}}
\label{wellposed_flux}
\\
\times \bigg\{ {} & \bigg( \mC_{k\kp} +  \mC_{\kp k} \bigg) 
\nonumber
\\
+ {} & \bigg( \mC_{k+\kp,-k} + \mC_{k+\kp,-\kp} \bigg)
\nonumber
\\
+ {} & \bigg( \gamma_{k\kp} \, \mC_{k,-k-\kp} + (1 \!-\! \gamma_{k\kp}) \, \omC_{k,-k-\kp} \bigg)
\nonumber
\\
+ {} & \bigg( \gamma_{k\kp} \mC_{\kp,-k-\kp} + (1 \!-\! \gamma_{k\kp}) \, \omC_{\kp,-k-\kp} \bigg)
\bigg\} ,
\nonumber
\end{align}
with the weight
\begin{equation}
\gamma_{k\kp} = \frac{(k \!-\! \kp)^{2}}{k^{2} \!+\! k^{\prime 2}} .
\label{def_gamma}
\end{equation}
Such a choice is always well-defined
since ${ k,\kp \!>\! 0 }$ (see Eq.~\ref{fund_res}).
This symmetrisation
is the same as in~\cite{Fouvry+2020}.

We can then rewrite Eq.~\eqref{wellposed_flux} as
\begin{equation}
\mF_{k\kp} (\Omega) = \mP \!\! \int \!\! \rd \Omega_{1} \, \frac{K_{k\kp} (\Omega , \Omega_{1})}{(\Omega \!-\! \Omega_{1})^{4}} ,
\label{rewrite_Flux}
\end{equation}
where we used the fact that ${ J \!\mapsto\! \Omega (J) }$
is monotonic, introducing therefore
${ \Omega \!=\! \Omega [J] }$ and ${ \Omega_{1} \!=\! \Omega[J_{1}] }$.
After calculation~\citep{MMA}, for ${ \delta \Omega \!\to\! 0 }$,
one finds
\begin{equation}
K (\Omega , \Omega \!+\! \delta \Omega) \simeq \mO \big[ \big( \delta \Omega \big)^{3} \big] ,
\label{DL_K}
\end{equation}
making the principal value
in Eq.~\eqref{rewrite_Flux} well-posed.

\section{Steady states}
\label{app:Steady}

\subsection{$H$-theorem}
\label{app:HTheorem}

Starting from Eq.~\eqref{def_S}, the time-derivative
of Boltzmann's entropy is given by
\begin{equation}
\frac{\rd S}{\rd t} = \!\! \int \!\! \rd J \, \frac{F'(J)}{F(J)} \, \mF (J) .
\label{calc_dSdt}
\end{equation}
Using the same symmetrisations as in Eq.~\eqref{final_calc_cons},
we find
\begin{align}
\frac{\rd S}{\rd t} = \third \sum_{\mathclap{k_{1} , k_{2}}} \!\! \int \!\! {} & \rd \bJ \, A_{k_{1}k_{2}} (\bJ)
\label{calc_dSdt_progress}
\\
\times {} & \bigg\{ (k_{1} \!+\! k_{2}) \frac{F'(J)}{F(J)} \!-\! k_{1} \frac{F'(J_{1})}{F(J_{1})} \!-\! k_{2} \frac{F'(J_{2})}{F(J_{2})} \bigg\} .
\nonumber
\end{align}
It then only remains to inject the expression
of ${ A_{k_{1}k_{2}} (\bJ) }$ (Eq.~\ref{sym_Flux_1})
to obtain ${ \rd S / \rd t }$ as given in Eq.~\eqref{dSdt}.

\subsection{Second-order kinetic blocking}
\label{app:2nd}

Let us consider interaction potentials
of the form
\begin{equation}
U_{k} (J , \Jp) \propto \frac{\delta_{k|d}}{|k|^{\alpha}} \, \big| J \!-\! \Jp \big|^{\alpha} ,
\label{generic_Uk}
\end{equation}
with $\alpha$ an arbitrary power law index
and ${ d \!\geq\! 1 }$ an arbitrary integer.
We also introduced the function
${ \delta_{k|d} \!=\! 1 }$ if ${ k \!\equiv\! 0 \,[\mathrm{mod}\,d] }$
and $0$ otherwise.

As detailed in~\cite{MMA}, for such a potential,
when embedded within the frequency profile ${ \Omega (J) \!\propto\! J }$,
one finds
\begin{align}
\mU_{k_{1}k_{2}} (J, J_{1} , J_{2}^{\res}) \propto \big[ {} & (k_{1} \!+\! k_{2}) \, \delta_{k_{1}|d} \, \delta_{k_{2}|d} 
\nonumber
\\
- {} & k_{1} \, \delta_{k_{1}+k_{2}|d} \, \delta_{k_{2}|d}
\nonumber
\\
- {} & k_{2} \, \delta_{k_{1}+k_{2}|d} \, \delta_{k_{1}|d} \big]^{2}
\nonumber
\\
= {} & 0 ,
\label{mU_blocked}
\end{align}
where we used
${ \delta_{k|d} \delta_{\kp|d} \!=\! \delta_{k|d}\delta_{k+\kp|d} }$.
We have therefore devised
a generic second-order kinetic blocking.

For ${ \alpha \!=\! 0 }$, Eq.~\eqref{generic_Uk}
becomes the simple interaction potential
${ U_{k} (J , \Jp) \!\propto\! \delta_{k|d} }$.
This makes the pairwise interaction independent
of the particles' actions.
This class of potential was already unveiled in~\cite{Fouvry+2020},
when investigating homogeneous ${1D}$ systems at order ${1/N^2}$.
In particular, \cite{Fouvry+2020} showed that this interaction potential becomes
\begin{equation}
U (\bw , \bwp) \propto 1 - \frac{1}{d} \sum_{k = 0}^{d-1} \deltaD \big[ (\theta \!-\! \thetap) \!-\! k \tfrac{\pi}{d} \big] .
\label{Uk_hom}
\end{equation}
As discussed in~\cite{Fouvry+2020},
such potentials amount to exactly local interactions:
they do not drive any relaxation.

For ${ d \!=\! 1 }$
and ${ \alpha \!>\! 0 }$ an even integer,
one can explicitly compute ${ U(\bw , \bwp) }$
from ${ U_{k} (J,\Jp) }$ as given by Eq.~\eqref{generic_Uk}.
Up to prefactors, one obtains~\citep[see 9.622.3 in][]{Gradshteyn+2007}
\begin{equation}
U (\bw , \bwp) \propto \big( J \!-\! \Jp \big)^{2n} B_{2n} \big[ \tfrac{1}{2 \pi} w_{2 \pi} (\theta \!-\! \thetap) \big] ,
\label{U_modelB}
\end{equation}
with ${ n \!\geq\! 1 }$ an integer.
In that expression, ${ B_{2n} (x) }$ is a Bernoulli polynomial,
e.g.\@,
${ B_{2} (x) \!=\! x^{2} \!-\! x \!+\! \tfrac{1}{6} }$.
We also introduced the ``wrapping'' function
\begin{equation}
w_{2\pi} (\theta) \equiv \theta \, [\mathrm{mod} \, 2 \pi] ;
\quad
0 \leq w_{2 \pi} (\theta) < 2 \pi .
\label{def_w2pi}
\end{equation}
Investigating systems driven by interaction potentials
as in Eq.~\eqref{U_modelB}
will be the subject of future works.

\section{Numerical simulations}
\label{app:NBody}

We consider the same system as in~\cite{Fouvry+2019},
i.e.\ a set of $N$ particles evolving on the unit sphere.
This setup mimics ``vector resonant relaxation"
in galactic nuclei~\citep{Kocsis+2015}.
Given an axis $\hbz$, the spherical coordinates are denoted with ${ (\vtheta,\phi) }$,
so that ${ \bw \!=\! [\phi \!=\! \theta, \cos (\vtheta) \!=\! J] }$
are canonical coordinates
with the associated unit vector, ${ \hbL \!=\! \hbL (\bw) }$.

Following~\cite{Barre+2014}, we fix the external potential,
$\Uext$, and the pairwise interaction potential, $U$,
to be
\begin{equation}
\Uext (\hbL) = D \, \big( \hbL \!\cdot\! \hbz \big)^{2} ;
\quad
U (\hbL , \hbLp) = G \, \hbL \!\cdot\! \hbLp .
\label{U_sphere}
\end{equation}
Following~\cite{Kocsis+2015}, the individual equations of motion read
${ \rd \hbL_{i} / \rd t \!=\! \p H / \p \hbL_{i} \!\times\! \hbL_{i} }$.
For the particular choice from Eq.~\eqref{U_sphere}, we readily find
\begin{equation}
\frac{\p H}{\p \hbL_{i}} = 2 D \, \big( \hbL_{i} \!\cdot\! \hbz \big) \, \hbz + G \, \bS ,
\label{calc_grad_H}
\end{equation}
with ${ \bS \!=\! \sum_{i = 1}^{N} \! \mu \, \hbL_{i} }$
the system's magnetisation.
Since $\bS$ is ``shared''
by all the particles, the rates of changes, ${ \{ \rd \hbL_{i} / \rd t \}_{i} }$,
can be computed in ${ \mO (N) }$ operations.

The invariants of the present system are
\begin{equation}
\forall i \!:\, \big| \hbL_{i} \big| = 1 ;
\quad
S_{z} = \bS \!\cdot\! \hbz ;
\quad
\Etot \!=\! H .
\label{invariants_sphere}
\end{equation}
Because of the gauge invariance associated with the constraints,
${ |\hbL_{i}| \!=\! 1 }$,
${ \rd \hbL_{i} / \rd t }$ is uniquely defined,
while the gradient
${ \p H / \p \hbL_{i} }$ is not.
As such, we define, unambiguously, the precession vector,
${ \bO_{i} \!=\! \hbL_{i} \!\times\! \rd \hbL_{i} / \rd t }$,
so that
\begin{equation}
\frac{\rd \hbL_{i}}{\rd t} = \bO_{i} \times \hbL_{i} ;
\quad
\bO_{i} \!\cdot\! \hbL_{i} = 0 .
\label{EOM_bO}
\end{equation}
In order to exactly conserve all the ${ |\hbL_{i}| }$,
the system's dynamics is integrated using the structure-preserving
classical fourth-order Munthe--Kaas scheme~\citep{MuntheKaas1999}.
We refer to~\S{5.1} in~\cite{Fouvry+2022}
for a presentation of this explicit scheme.

For the results presented in Fig.~\ref{fig:Flux},
we considered ${ N \!=\! 1\,024 }$, ${ G \!=\! -1 }$,
${ D \!=\! 15 }$, ${ \Mtot \!=\! 1 }$.
The integration was performed with a fixed timestep
${ h \!=\! 5 \!\times\! 10^{-4} }$, with a dump every ${\Delta t \!=\! 100}$
and integrated up to ${ t \!=\! 2 \!\times\! 10^{5} }$.
Running one realisation requires about ${32\,\mathrm{h}}$
of computation on a single core.
At the final times, the relative errors
in $S_{z}$ and $\Etot$ were typically $10^{-7}$.

For the initial conditions, as in~\cite{Fouvry+2019},
we consider
\begin{equation}
F (\bw) = C \, \re^{- (J/\sigma)^{4}} ,
\label{choice_F_sphere}   
\end{equation}
with $C$ a normalisation constant.
We fixed the \DF\@'s initial dispersion to
${ \sigma \!=\! 0.45 }$.
This ensures that the system
is linearly stable and hot enough~\citep[see fig.~{6} in][]{Fouvry+2019}.

To obtain Fig.~\ref{fig:Flux},
we performed a total of ${ 1\,024 }$ independent realisations.
The action, ${ -1 \!\leq\! J \!\leq\! 1 }$, is binned in $50$
equal size bins.
For each realisation, each action bin and each time dump,
we determine the number of particles left to that action.
This is then averaged over the available realisations.
Finally, we fit the associated time series
with an affine time dependence: its slope
is proportional to the local diffusion flux, ${ \mF (J , t \!=\! 0) }$.
To estimate the measurement errors, we follow
the same bootstrap approach as in~\S{F} of~\cite{Fouvry+2019N2}.
In Fig.~\ref{fig:Flux}, we report the 16\% and 84\% level lines
over $100$ bootstrap resamplings.

\end{document}